\documentclass[twocolumn,breaklinks=true]{aastex631}

\usepackage{natbib}
\usepackage{multirow}
\usepackage{gensymb} % needed for the \degree symbol
\usepackage[flushleft]{threeparttable} 
\usepackage{amsmath}
\usepackage{amssymb}
\usepackage{url}
\usepackage{pifont}
\usepackage{wasysym}

\setcitestyle{comma}

\shorttitle{Young stellar populations ages at $z\sim10$ from IRAC}
\shortauthors{Stefanon et al.}

\begin{document}

\title{Deep Spitzer/IRAC Data for $z\sim10$ galaxies Reveal Blue Balmer Break Colors: Young Stellar Populations at $\sim500$ Myr of Cosmic Time }

\author{Mauro Stefanon}
\affiliation{Departament d'Astronomia i Astrof\`isica, Universitat de Val\`encia, C. Dr. Moliner 50, E-46100 Burjassot, Val\`encia,  Spain}
\affiliation{Observatori Astron\`omic, Universitat de Val\`encia, C/ Catedr\'atico Jos\'e Beltran, 2, 46980 Paterna Val\`encia, Spain}
\affiliation{Unidad Asociada CSIC "Grupo de Astrof\'isica Extragal\'actica y Cosmolog\'ia" (Instituto de F\'isica de Cantabria - Universitat de Val\`encia)}
\affiliation{Leiden Observatory, Leiden University, NL-2300 RA Leiden, Netherlands}

\author{Rychard J. Bouwens}
\affiliation{Leiden Observatory, Leiden University, NL-2300 RA Leiden, Netherlands}

\author{Ivo Labb\'e}
\affiliation{Centre for Astrophysics and SuperComputing, Swinburne, University of Technology, Hawthorn, Victoria, 3122, Australia}

\author{Garth D. Illingworth}
\affiliation{UCO/Lick Observatory, University of California, Santa Cruz, 1156 High St, Santa Cruz, CA 95064, USA}

\author{Valentino Gonzalez}
\affiliation{Departamento de Astronom\'ia, Universidad de Chile, Casilla 36-D, Santiago 7591245, Chile}
\affiliation{Centro de Astrof\'isica y Tecnologias Afines (CATA), Camino del Observatorio 1515, Las Condes, Santiago 7591245, Chile}

\author{Pascal A. Oesch}
\affiliation{Departement d'Astronomie, Universit\'e de Gen\'eve, 51 Ch. des Maillettes, CH-1290 Versoix, Switzerland}
\affiliation{Cosmic Dawn Center (DAWN), Niels Bohr Institute, University of Copenhagen, Jagtvej 128, K\o benhavn N, DK-2200, Denmark}

\email{Email: mauro.stefanon@uv.es}

\begin{abstract}

We present the deepest constraints yet on the median rest-UV+optical SED of $z\sim10$ galaxies, prior to JWST science operations.  We constructed stacks based on four robust $J_{125}$-dropouts, previously identified across the GOODS fields. We used  archival HST/WFC3 data and the full depth Spitzer/IRAC mosaics from the GREATS program, the deepest coverage at $\sim3-5\mu$m to date. The most remarkable feature of the SED is a blue IRAC $[3.6]-[4.5]=-0.18\pm0.25$ mag color. We also find a nearly flat $H_{160}-[3.6]=0.07\pm0.22$ mag color, corresponding to a UV slope $\beta= -1.92\pm0.25$. This is consistent with previous studies, and indicative of minimal dust absorption. The observed blue IRAC color and SED fitting suggest that $z\sim10$ galaxies have very young (few $\times10$ Myr) stellar populations, with $80\%$ of stars being formed in the last $\lesssim 160$ Myr ($2\sigma$). While an exciting result, the uncertainties on the SED are too large to allow us to place strong constraints on the presence of a nebular continuum in $z\sim10$ galaxies (as might be suggested by the blue $[3.6]-[4.5] < 0$ mag color). The resulting sSFR is consistent with the specific accretion rate of dark matter halos, indicative of a star-formation efficiency showing quite limited evolution at such early epochs. 

\end{abstract}


\section{Introduction}
\label{sect:intro}

Understanding how efficiently stars formed out of the cold gas accreted through the potential wells of the hierarchical assembly of the dark matter halos is one of the most fundamental questions in modern astrophysics (e.g., \citealt{madau2014}).

Over the last decade, the sensitive and high spatial resolution data acquired by the Wide Field Camera 3 (WFC3) onboard the Hubble Space Telescope (HST) has enabled the detection and analysis of the rest-frame UV light emitted by recently-born massive O-B stars (e.g., \citealt{kennicutt1998}) up to $z\sim10-12$ (\citealt{bouwens2013, coe2013, oesch2014, bouwens2015, mcleod2016, oesch2016, calvi2016, bernard2016, oesch2018, salmon2018, ishigaki2018, morishita2018, bouwens2019a, lam2019, salmon2020, strait2020, bouwens2022, finkelstein2022} - see also \citealt{harikane2022} for $z\sim12$ candidates identified from ground-based NIR data). These studies resulted in fundamental constraints to the total budget of newly-formed stars across $\gtrsim13$ Gyr of lookback time (the so-called cosmic star-formation rate density - CSFRD - e.g., \citealt{bouwens2008,oesch2014, bouwens2015, finkelstein2015a, bouwens2016, mcleod2016, oesch2018, ishigaki2018, bhatawdekar2019, bouwens2022}).

Nonetheless, the highly successful Spitzer/IRAC camera (\citealt{fazio2004}) has provided the community with an invaluable probe into the rest-frame optical light for galaxies at $z>4$, complementing the information in the rest-frame UV from HST. These data have unveiled a very active Universe, where high equivalent widths EW ($>500-1000$\AA) line emission from the [\ion{O}{3}]+H$\beta$ (
\citealt{labbe2013, smit2014, castellano2017, debarros2019, stefanon2019, bowler2020, strait2020, strait2021, endsley2021b, stefanon2022a}) and H$\alpha$  (\citealt{smit2016, bouwens2016c, rasappu2016, faisst2016, marmol-queralto2016, caputi2017, harikane2018b, lam2019, faisst2019, maseda2020, stefanon2022b}) are typical, and allowed to probe the stellar masses of galaxies at $z\sim 4-9$ (\citealt{duncan2014, grazian2015, song2016, bhatawdekar2019, kikuchihara2020}), up to $z\sim10$ (\citealt{oesch2016, stefanon2021a}).

The enormous progress in constraining the CSFRD and stellar mass, however, has not resulted yet in a clear consensus on the star-formation efficiency in the early Universe. For example, some studies report a  $\sim10\times$ higher abundance of galaxies at $6<z<10$ (e.g., \citealt{ellis2013, mcleod2016}), and older stellar populations (\citealt{hashimoto2018, roberts-borsani2020, mawatari2020, tacchella2022}), qualitatively consistent with a star-formation efficiency increasing with increasing redshifts at early epochs (e.g., \citealt{finkelstein2015b, behroozi2019}). However, a number of more recent determinations of the star-formation rate and stellar mass densities using the largest samples suggest that the baryonic assembly is consistent with the accretion of the dark matter halos (\citealt{oesch2018, tacchella2018, stefanon2021a}). Clearly, JWST will rather quickly establish  if the latest results are the most likely (or not!).

Interestingly, but perhaps not surprisingly, the current redshift frontier ($z\sim9-12$) is where systematic differences in the stellar mass and stellar-to-halo mass ratio determinations are the most evident (see e.g., \citealt{stefanon2021a}). Again, in this context, JWST observations will provide much more extensive insights, but we have the opportunity to make a significant step forward on this important question with existing data.

Observationally, the $M_\star/L$ ratio (where $M_\star$ indicates the stellar mass, and $L$ the luminosity in a specific band), or, equivalently, the age of the stellar population, is arguably one of the main ingredients required to constrain the stellar mass assembly of galaxies. The Balmer/$4000$ \AA\ break, one of the main proxies for stellar population age determinations (e.g., \citealt{kauffmann2003}), is probed by the IRAC $3.6\mu$m band for redshifts in the range $7.2\lesssim z \lesssim8.7$.  At $z\sim8.7$ the Balmer/4000\AA\ break starts to leave the IRAC $3.6\mu$m band, and by $z\sim9.7$ it has completely shifted into the $4.5\mu$m band. The $4.5\mu$m band, instead, at $7.0\lesssim z \lesssim 9.2$ intercepts the emission from the [\ion{O}{3}]$_{\lambda\lambda 4959,5007}$ and H$\beta$ lines, while for $z\gtrsim 9.2$ it is largely free of the most prominent emission lines (e.g., \citealt{anders2003}).

Fortunately, we can constrain the Balmer/$4000$\AA\ break for $z\sim10$ galaxies by leveraging Spitzer/IRAC observations. At these redshifts, the IRAC $3.6\mu$m band probes the rest-frame UV ($\sim 3300$\AA\ rest-frame), while the $4.5\mu$m band lies just red-ward of the Balmer/$4000$ \AA\ break. This observational configuration is particularly effective at constraining the strength of the Balmer/4000\AA\ break given the contiguity of the two IRAC bands. Moreover, at $z\lesssim9$ the typical uncertainties in photometric redshifts ($\Delta z\sim0.4$) can prevent from distinguishing genuine Balmer/$4000$ \AA\ break from emission line contributing the $4.5\mu$m band  (see e.g., Figure 5 in \citealt{stefanon2021a}), leading to potential overestimates of the stellar age (e.g., \citealt{stark2013, stefanon2021a, topping2022}). Instead, the larger shift to redder wavelengths and the absence of strong line emission just red-ward of the Balmer break make this possibility less likely for $z\sim10$ sources.

Thus, given the value of $z\sim10$ samples for constraining the SFH, we focus on inferring the main stellar population parameters for a robust sample of star-forming galaxies at $z\sim10$ (\citealt{oesch2018}). In particular, we characterize the rest-frame optical light, complementing the previous study of \citet{wilkins2016} centered on the rest-frame UV light. We do so by stacking the very sensitive observations available with Spitzer/IRAC data over the GOODS fields (\citealt{giavalisco2004}) from the GOODS Re-ionization Era wide Area Treasury from Spitzer (GREATS) + other programs (\citealt{stefanon2021b}). GREATS combined all the relevant IRAC data ($> 4200$ hr cumulative  in the $3.6$ and $4.5\mu$m bands) acquired throughout the scientific life of Spitzer, providing us with the deepest available $3 - 5\mu$m observations prior to the start of JWST science operations.

A brief description of the layout of this paper follows. In Section \ref{sect:sample} we introduce the adopted sample and describe the relevant datasets, while in Section \ref{sect:analysis} we summarize our stacking procedure and the configurations we adopted to infer the main physical parameters. The resulting median spectral energy distribution (SED) is presented in Section \ref{sect:results}, while in Section \ref{sect:discuss} we discuss its implications in galaxy assembly in the early Universe. We summarize our findings in Section \ref{sect:conclusions}.

Throughout this paper, we adopt $\Omega_M=0.3$, $\Omega_\Lambda=0.7$ and $H_0=70$\,km s$^{-1}$ Mpc$^{-1}$, consistent with the most recent estimates from Planck (\citealt{planck2020_cosmo}). Magnitudes are given in the AB system (\citealt{oke1983}), while our $M_\star$ and SFR measurements are expressed in terms of the \citet{salpeter1955} initial mass function (IMF). For brevity, we denote the \textit{HST} F435W, F606W, F775W, F850LP, F105W, F125W, F140W and F160W as $B_{435}$, $V_{606}$, $i_{775}$, $z_{850}$, $Y_{105}$, $J_{125}$, $JH_{140}$ and $H_{160}$, respectively.

\begin{figure}
\begin{tabular}{cc}
\hspace{-1cm}\includegraphics[width=9cm]{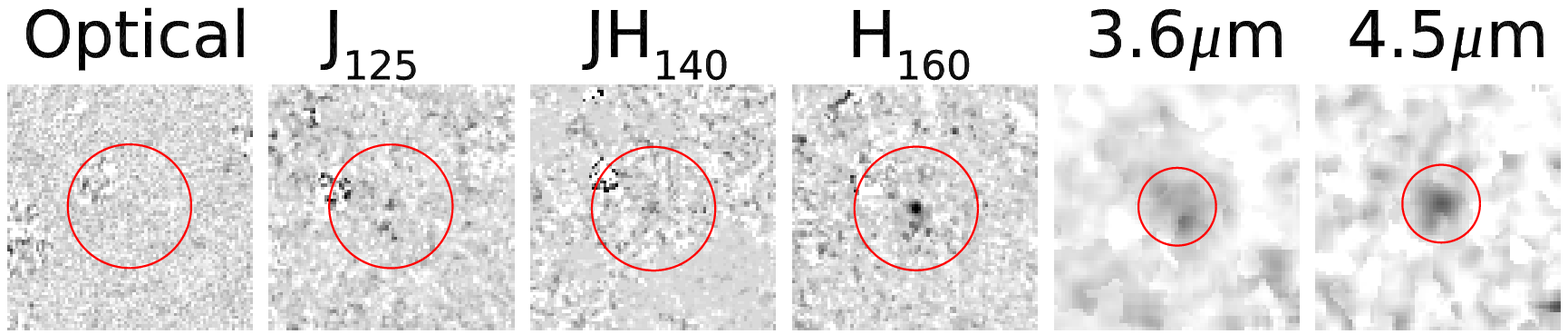} \\
\hspace{-1cm}\includegraphics[width=9cm]{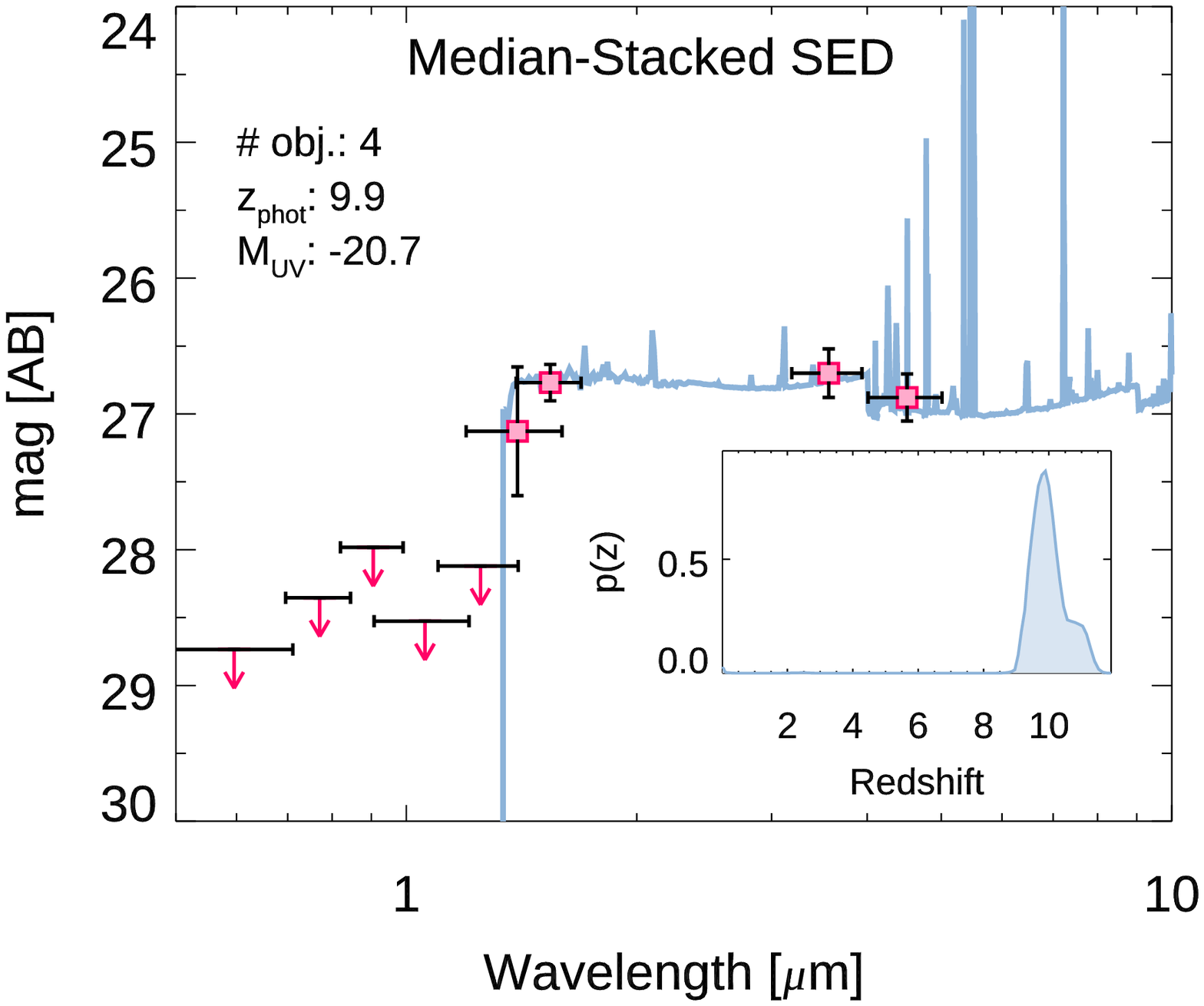}
\end{tabular}
\caption{\textit{Top panels:} Stacked image stamps in the \textit{HST}  and IRAC bands, as labelled at the top, centered on the median stacks. The stamps for the \textit{HST} bands are $5\farcs0$ per side, while those in the IRAC bands are $8\farcs0$ wide. The \textit{HST} optical stack combines all the data available blue-ward of the $J_{125}$ band (i.e., $B_{435}, V_{606}, i_{775}$ and $z_{850}$). The red circle marks the $2\farcs5$ diameter aperture adopted for our Spitzer/IRAC photometry.  The \textit{HST} stacks are shown for context, but we emphasise that our median \textit{HST} flux measurements were derived from our photometry on individual sources. \textit{Bottom panel:} Median-stacked SED resulting from our analysis. The stacked photometry is presented both as filled red squares with $1\sigma$ errors or as red downward-pointing arrows indicating $2\sigma$ upper limits. The effective width of each band is marked by the black horizontal bar.  The blue curve corresponds to the best-fitting \textsc{EAzY} template. The inset shows the redshift probability distribution computed by \textsc{EAzY}. Labels at the top-left corner indicate the number of objects entering the stack, the median redshift, and the $M_\mathrm{UV}$ computed by $\textsc{EAzY}$ (top to bottom, respectively). The SED is characterized by a blue IRAC $[3.6]-[4.5]$ color, indicative of young stellar populations.  \label{fig:SED}}
\end{figure}

\begin{deluxetable}{lcccc}
\tablecaption{Sample of $z\sim10$ star-forming galaxies adopted in this work. \label{tab:sample}}
\tablehead{\colhead{ID} & \colhead{Redshift} & \colhead{$M_\mathrm{UV}$}  & \multicolumn{2}{c}{$5\sigma$ sensitivities\tablenotemark{$\ddagger$}}   \\
 & & & \colhead{$3.6\mu$m} & \colhead{$4.5\mu$m} \\
 & & \colhead{(mag)}  & \colhead{(nJy)} & \colhead{(nJy)}   %\\
}
\startdata
GN-z11    & $11.1\pm0.1$\tablenotemark{*} & $-21.6$ & 53 & 78  \\
GN-z10-2 & $9.9\pm0.3$ & $-20.7$ &  67 & 103  \\
GS-z10-1 & $9.9\pm0.5$ & $-20.6$ &  51 & 79 \\
XDFJ-3811362435 & $9.9^{+0.7}_{-0.6}$ & $-17.6$ &  48 & 77\\
\hline
\multicolumn{5}{c}{Excluded\tablenotemark{$\dagger$}} \\
GN-z10-3\tablenotemark{a} & $8.78$ & $-20.7$ & 71 & 124 \\
XDFJ-4023680031\tablenotemark{b} & $9.7\pm0.6$ & $-17.4$ & 49 & 79 \\
\enddata
%}
\tablenotetext{*}{Spectroscopic redshift from \citet{oesch2018}. \citet{jiang2021} find a spectroscopic redshift of $10.957\pm0.001$ for this source.}
\tablenotetext{\ddagger}{Nominal $5\sigma$ sensitivities from the IRAC SENS-PET exposure time calculator adopting the coverage depths from the GREATS mosaics.}
\tablenotetext{\dagger}{Sources in the GOODS fields listed in \citet{oesch2018} or \citet{bouwens2015}  but excluded from our analysis.}
\tablenotetext{a}{This source appears as a bona-fide $J$-dropout in the selections of \citet{oesch2014, oesch2018}; however, given reports of Lyman-$\alpha$ emission from the source and its thus seeming to lie at $z = 8.78$ (\citealt{laporte2021}), we exclude it from our stack results.}
\tablenotetext{b}{Removed from our sample given the significant residuals present in the IRAC data after subtraction of the neighboring sources.}
\end{deluxetable}

\section{Data}
\label{sect:sample}

The initial $z\sim10$ sample considered for this study is composed of the five candidate  $z\sim10$ Lyman-break galaxies identified as $J_{125}$-dropouts  by \citet{oesch2013b,oesch2014,oesch2016,oesch2018, bouwens2010b,bouwens2011, bouwens2015, bouwens2016} across the CANDELS (\citealt{grogin2011, koekemoer2011}) GOODS-N, GOODS-S (\citealt{giavalisco2004}), the ERS field (\citealt{windhorst2011}), and the UDF/XDF (\citealt{beckwith2006,illingworth2013, ellis2013}) with the HUDF09-1 and HUFD09-2 parallels (\citealt{bouwens2011b}). We complemented this sample with GN-z11 (\citealt{oesch2016}), a star-forming galaxy spectroscopically confirmed at $z_\mathrm{spec}=11.1$ (\citealt{oesch2016} - see also \citealt{jiang2021}). In Table \ref{tab:sample} we list the sources considered for this study and the main properties of the imaging data sets. The HST mosaics are characterized by $5 \sigma$ depths of $\approx27.5$~mag in the $V_{606}$ and $I_{814}$ bands,  $\approx26.7-27.5$~mag in the $Y_{105}$, and $\sim26.8-27.8$\,mag in the $J_{125}$ and $H_{160}$ bands. An essential aspect for our study is the availability of exquisitely-deep coverage at $3-5\mu$m in the  GOODS fields from the Spitzer/IRAC GREATS program (PI: Labb\'e - \citealt{stefanon2021b}). These mosaics combine all the useful IRAC imaging data acquired during the $>15$ years of Spitzer science operations, with typical $5 \sigma$ depths of  $\sim26.0-27.0$\,mag in the IRAC $3.6\mu$m and $4.5\mu$m bands. 

Considering the typical uncertainties in photometric redshifts ($\Delta z\lesssim 0.6$), our $z\sim10$ sample selection minimizes the potential contamination of line emission which could affect the $4.5\mu$m band for $z\lesssim9$ sources (see e.g., Figure~5 of \citealt{stefanon2021a}), increasing the robustness of the inferred physical properties.

To overcome the challenge posed by the broad Spitzer/IRAC PSF and light from neighbouring sources contaminating flux measurements at $>3 \mu$m, we followed the procedure used in \citet{stefanon2021a,stefanon2022a} and \citet{stefanon2022b} and subtracted the light from neighbouring sources with \textsc{Mophongo} \citep{labbe2006, labbe2010a, labbe2010b, labbe2013, labbe2015}. A critical ingredient for this step was the availability of accurate location-dependent PSFs from GREATS, as the asymmetric profile of the instrumental IRAC PSF and the variety of programs included in the mosaics result in significant variation of the PSF light profile across the mosaics (see e.g., \citealt{stefanon2021b}). As a result of this process,  one source (XDFJ-4023680031) was removed from our sample as it showed residual contamination after visual inspection.

Finally, we removed GN-z10-3 to be conservative.   While it was initially identified at having a redshift of  $z\sim10$ (\citealt{oesch2014}), \citet{laporte2021} report the detection of a candidate Ly$\alpha$ line, suggesting the redshift may be $z_{\mathrm{Ly}\alpha}=8.78$  (\citealt{laporte2021}).  Our analysis therefore was conducted without GN-z10-3 and on the remaining four sources (see Table \ref{tab:sample}).

\section{Analysis}
\label{sect:analysis}

We applied our consolidated stacking procedure (\citealt{stefanon2022a, stefanon2022b} for details) to construct the median spectral energy distribution (SED) of $z\sim10$ star-forming galaxies. We median-stacked the photometry in the HST bands normalized by the flux density in the $H_{160}$ band of each source. Uncertainties associated with the flux densities were computed by bootstrapping the procedure $1000$ times. The stacked flux density in the $H_{160}$ band instead corresponds to the median of the individual photometric measurements, while the associated uncertainty was computed by bootstrapping the procedure $1000$ times. For the IRAC $3.6\mu$m and $4.5\mu$m bands, instead, we median-stacked the image stamps centered on each source after normalizing them by the corresponding $H_\mathrm{160}$ flux density. For this step we combined the stamps obtained after removing the contribution of all neighbouring sources within a radius of $9\farcs0$. The flux density in the stacked IRAC stamps was measured in apertures of $2\farcs5$ diameter, and corrected to total flux densities using the median of the PSFs reconstructed at the location of each source (the applied correction factors are $1.85$ and $1.91$, for the $3.6$ and $4.5\mu$m band, respectively). Uncertainties were computed by measuring the dispersion of the flux density estimates in $20$ $2\farcs5$-wide apertures randomly placed across the stacked stamp within $7.5\farcs$ of the center, and repeating the process $10\times$ to increase the statistical significance. Finally, all values were rescaled by the median of the flux density in the $H_{160}$ band. These stacked images then were used in our stellar population analysis.\\

We estimated the main stellar population parameters using the bayesian tool \textsc{Prospector} \citep{johnson2021}, which runs on the \textsc{Flexible Stellar Population Synthesis (FSPS)} package (\citealt{conroy2009, conroy2010}) with the \textsc{Modules for Experiments in Stellar Astrophysics Isochrones and Stellar Tracks} (MIST; \citealt{choi2016, dotter2016}). Our estimates are based on a \citet{salpeter1955} IMF defined between $0.1$ and $240 M_\odot$, a \citet{calzetti2000} extinction curve, a  $Z_\mathrm{star}\equiv Z_\mathrm{gas}=0.2Z_\odot$ metallicity, a ionization parameter  $\log U = -2.5$  (e.g., \citealt{stark2017, debarros2019}), and a formation redshift of $z=20$ (\citealt{mawatari2020,hashimoto2018, tacchella2022, harikane2022}). 

Given that recent studies indicate that parametric star-formation histories (SFHs) could be underestimating the masses at high redshifts (e.g., \citealt{leja2019b, topping2022}), we considered two different star formation histories: a constant (CSFH), and a non-parametric SFH. In particular, we assumed that the non-parametric SFH was defined by four bins in lookback time, which we fixed at $0-3$Myr, $3-13$Myr, $13-100$ Myr, and $100-300$ Myr, respectively, with a Student's t-distribution continuity prior modulating the ratio of the SFR in contiguous bins, with $\nu=2$ and $\sigma=2$ (see discussions in \citealt{leja2019a, tacchella2022}). This configuration can concurrently accommodate  a recent burst of star formation and significant star formation during the initial assembly of the galaxy (\citealt{hashimoto2018,roberts-borsani2020, tacchella2022}). We also repeated the process with up to 8 bins in lookback time ($0-3$Myr, $3-13$Myr, $13-100$ Myr, and log-spaced at later lookback times) and found similar results. 

Finally, we also considered a third model with a constant SFH and pure stellar emission (i.e., assuming the nebular continuum and line emission are negligible). During the fits, the redshift was fixed to the median of the photometric redshifts of the sample ($z_\mathrm{median}$). Instead, the SFR integrated over the cosmic time (i.e, the total mass $M_\mathrm{T}$) and the dust optical depth at $5500$\AA\ ($\tau_{5500}$) varied under a flat prior ($6\le\log M_\mathrm{T}/M_\odot\le 11$ and $0\le \tau_{5500} \le 5$, respectively). Nonetheless, the mass values quoted in this paper refer to the mass in surviving stars (i.e, stellar mass $M_\star$), obtained by rescaling $M_\mathrm{T}$ by the estimated stellar-to-total mass ratio (the \textsc{mfrac} parameter, with  typical values in the range $\sim0.8-1.0$).

SFR were computed by converting the UV luminosity using the factors listed by \citet{madau2014}, interpolated for a $Z=0.2Z_\odot$ metallicity and a \citet{salpeter1955} IMF. Because of the increasing indication of negligible dust emission at early epochs (e.g., \citealt{bouwens2016, dunlop2017, mclure2018, bouwens2020}), and considering the small $A_V$ values obtained with the non-parametric and CSFH configurations (see Section \ref{sect:discuss}), we opted for not correcting the SFR for dust extinction. We finally computed the sSFR ($\equiv$SFR/$M_\star$) combining the SFR and $M_\star$ values previously estimated.

\section{Results}
\label{sect:results}

\begin{deluxetable*}{lccccccccc}
\tablecaption{Flux densities for our median-stacked photometry \label{tab:stack_phot}}
\tablehead{ & \colhead{$V_{606}$} & \colhead{$i_{775}$} & \colhead{$z_{850}$}  & \colhead{$Y_{105}$} & \colhead{$J_{125}$} & \colhead{$JH_{140}$} & \colhead{$H_{160}$}  & \colhead{$3.6\mu$m} & \colhead{$4.5\mu$m}  \\
 & \colhead{(nJy)}  & \colhead{(nJy)} & \colhead{(nJy)} & \colhead{(nJy)} & \colhead{(nJy)} & \colhead{(nJy)}   & \colhead{(nJy)} & \colhead{(nJy)} & \colhead{(nJy)}  %\\
}
\startdata
Stack & $-1.0 \pm  5.8$ & $-0.6 \pm  8.3$ & $-5.1 \pm 11.6$ & $-4.7 \pm  7.1$ & $14.7 \pm 10.3$ & $51.2 \pm 22.3$ & $71.2 \pm  8.8$ & $75.8 \pm 12.5$ & $64.3 \pm 10.3$ \\
\enddata
%}
\tablecomments{We only list the flux densities in those bands available for at least $90\%$ of the sources in our sample. \label{tab:phot}}
\end{deluxetable*}

Table \ref{tab:phot} lists the photometric measurements from our stacking procedure, while in Figure \ref{fig:SED} we display the stacked stamps and the measured broad-band SED. To further validate our stacking analysis, and as an initial  guide to the interpretation of our results, in Figure \ref{fig:SED} we also present the best-fitting template from \textsc{EAzY} (\citealt{brammer2008} - see \citealt{stefanon2021a, stefanon2022a} for details on the template set we adopted with \textsc{EAzY}).

The most prominent feature consists of a blue $[3.6]-[4.5] = -0.18\pm0.25$ mag color. Because at $z\sim10$ the IRAC $3.6\mu$m band probes the rest-UV, while the $4.5\mu$m band lies just red-ward of the Balmer/$4000$\AA\ break, the $[3.6]-[4.5]$ color brackets the Balmer/$4000$\AA\ break, an indicator of stellar population age (e.g., \citealt{kauffmann2003}). The measured blue color then suggests young stellar populations (e.g., \citealt{inoue2011}), either as the result of a very rapid assembly of stars (e.g., an exponentially rising SFH) or as a recent burst of star formation. We will return to the implications of the $[3.6]-[4.5]$ color on the stellar populations at $z\sim10$ in Section \ref{sect:age}.

Our stacked SED is also characterized by an approximately flat $H_{160}-[3.6]=0.07\pm0.22$ mag color, corresponding to a UV slope $\beta=-1.92\pm 0.25$, marginally redder than, but consistent with, the previous estimates of $\beta=-2.1\pm0.3$ at $z\sim 10$ by \citet{wilkins2016a}. This result is not very surprising, considering that 3/4 of the sources in our sample are in common with that of \citet{wilkins2016a}. Our UV slope estimate is also consistent with the recent measurements for $z\sim9$ sources by \citet[$\beta=-2.1\pm0.4$ for $M_\mathrm{UV}\sim-21$ mag]{bhatawdekar2021} and \citet[$\beta=-1.9\pm0.3$]{tacchella2022}. It is reassuring to find consistent measurements, considering the IRAC data adopted for this work (\citealt{stefanon2021b}) have on average $\sim2\times$ longer exposure times than the mosaics available to \citet{wilkins2016a}.

\begin{deluxetable}{rccc}
\tablecaption{Main stellar population parameters inferred for our stacked SED \label{tab:pop_params}}
\tablehead{ \colhead{Property} & \multicolumn{3}{c}{Value}  
   %\\
}
\startdata
       $z_\mathrm{median}$ & \multicolumn{3}{c}{$  9.9$} \\
        $M_\mathrm{UV}$ & \multicolumn{3}{c}{$ -20.7\pm 0.1 $ mag} \\
       UV slope $\beta$ & \multicolumn{3}{c}{$  -1.92\pm 0.25 $} \\
      SFR$_\mathrm{UV}$\tablenotemark{$\dagger$} & \multicolumn{3}{c}{$   9.1_{-0.8}^{+0.9} M_\odot/\mathrm{yr}$} \\
SFH & Non Param. & CSFH & CSFH no neb. \\
          log(t$_{\ell,50}$/yr)\tablenotemark{$\ddagger$} & $   6.4_{ -0.2}^{ +1.2} $ & $   6.4_{ -0.4}^{ +0.7} $ & $   6.1_{ -0.4}^{ +0.2} $ \\ 
          log(t$_{\ell,80}$/yr)\tablenotemark{$\ddagger$} & $   7.6_{ -1.2}^{ +0.6} $ & $   6.6_{ -0.4}^{ +0.7} $ & $   6.3_{ -0.4}^{ +0.2} $ \\ 
$\log(M_\star/M_\odot)$ & $   8.4_{-0.2}^{+0.4} $ & $   8.4_{-0.1}^{+0.3} $ & $   9.3_{-0.4}^{+0.1} $ \\
                  $A_V$/mag & $   0.2_{-0.1}^{+0.1} $ & $   0.2_{-0.1}^{+0.1} $  & $   1.1_{-0.6}^{+1.1} $  \\
                   sSFR/Gyr$^{-1}$\tablenotemark{*} & $  33.4_{-19.8}^{+24.1} $  & $  40.0_{-36.0}^{+27.5} $  & $   4.6_{ -0.7}^{ +5.9} $  \tablenotemark{\eighthnote} \\
\enddata
\tablecomments{The measurements for $\log(t_{\ell,50}/\mathrm{yr})$, $\log(t_{\ell,80}/\mathrm{yr})$, $\log(M_\star/M_\odot)$, $A_V$ and sSFR refer to the median and $16-84$ percentiles of the posteriors.}
\tablenotetext{\dagger} {SFR computed from the rest-frame UV luminosity and assuming negligible dust obscuration. We warn the reader that, given the high fraction of obscured star formation in the configuration without nebular emission (CSFH no neb.), the corresponding SFR$_\mathrm{UV}$ value is potentially underestimated. Assuming a \citet{calzetti2000} curve, the unobscured+obscured SFR for this configuration would result in a $\approx12\times$ larger value. This however is not the option that we consider to be most likely.}
\tablenotetext{\ddagger}{Look-back time encapsulating $50\%$ and $80\%$ (log(t$_{\ell,50}$/yr) and log(t$_{\ell,80}$/yr), respectively) of stellar mass assembly.}
\tablenotetext{*}{sSFR obtained dividing the unobscured SFR$_\mathrm{UV}$ by the stellar mass $M_\star$.}
\tablenotetext{\eighthnote}{This value of sSFR does not account for the obscured fraction of SFR, and it is therefore likely underestimated in this case. A correction assuming a \citet{calzetti2000} extinction curve would result in a $\approx12\times$ larger value. Again this is not the option that we consider to be the most likely.}
%}
\end{deluxetable}

\begin{figure*}
\centering\includegraphics[width=18cm]{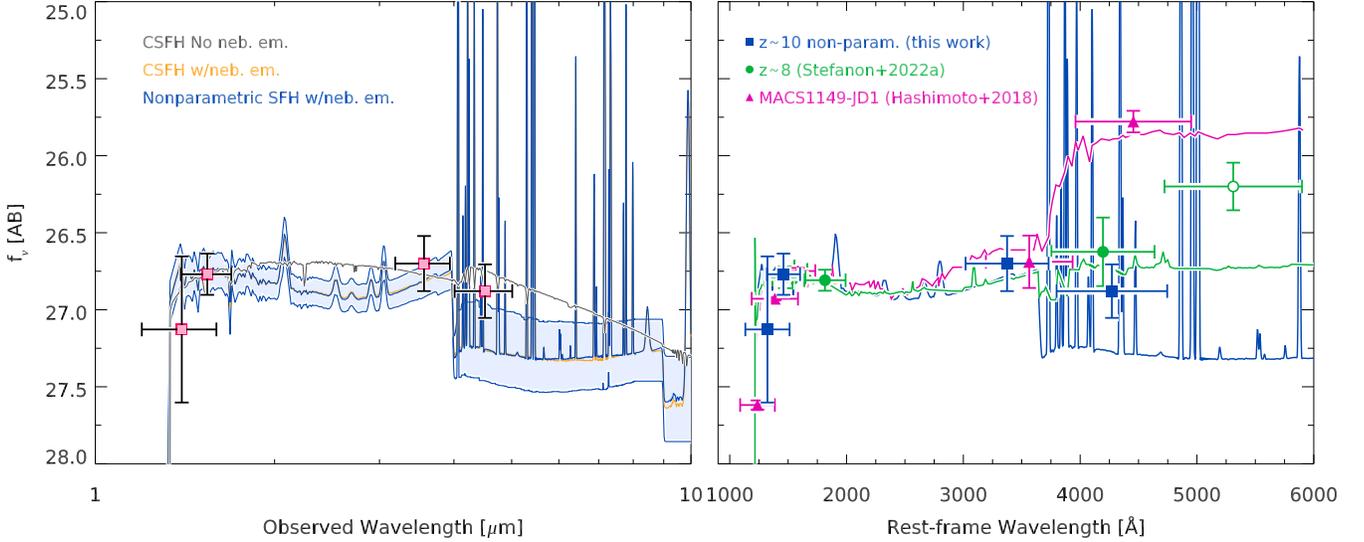}
\caption{[\textit{Left}] Best-fit SEDs resulting from the different configurations adopted in estimating the main stellar population parameters of our stack. In particular, we considered a constant SFH (CSFH) model with and without nebular emission (orange and grey solid curves, respectively), and a non-parametric SFH (blue curve - see Section \ref{sect:analysis} for details), which we adopt as reference (the orange and blue curves are essentially indistinguishable). The filled blue area corresponds to the SED posteriors marginalized over the $68\%$ confidence interval for the non-parametric configuration. [\textit{Right}] Comparison of our SED at $z\sim10$ (filled blue squares and curve) with results from other studies. Specifically, we show the median SED at $z\sim8$ corresponding to the $M_\mathrm{UV}\sim-20.7$ mag bin from \citet[green circles and curve]{stefanon2022a}, and the SED of MACS1149-JD1 \citep[filled purple triangles and curve - here we adopt the photometric measurements of \citealt{zheng2017}]{hashimoto2018}, a star-forming galaxy at $z_\mathrm{spec}=9.1$ characterized by an apparent, pronounced Balmer break (\citealt{hashimoto2018}). All wavelengths have been shifted to their rest frame. The photometric measurements and best-fit SEDs from \citet{stefanon2022a}, \citet{hashimoto2018} and \citet{zheng2017} have been renormalized to match the luminosity of our $z\sim10$ stack at $1600$\AA. We removed the main emission lines from the SEDs of \citet{stefanon2022a} and \citet{hashimoto2018} to improve readability, but note the likely substantial contribution of the [\ion{O}{3}]$_{\lambda\lambda4959,5007}$ doublet to the measured flux at $\lambda\sim5200$\AA\ for the \citet{stefanon2022a} results (open green circle). Our stacked photometry can be better described by young stellar population templates, a result likely driven by the observed blue $[3.6]-[4.5]\lesssim0$ mag color, consistent with the previous median determinations at $z\sim8$ by \citet{stefanon2022a}.  \label{fig:SEDs}}
\end{figure*}

\section{Discussion}
\label{sect:discuss}

Figure \ref{fig:SEDs} displays the results of our template fitting analysis, while in Table  \ref{tab:pop_params} we list the the values of the main physical parameters. In what follows, we adopt the results from the non-parametric SFH as reference. 

\subsection{Which are the typical stellar population ages at $\sim500$ Myr of cosmic time?}
\label{sect:age}

Interestingly, inspection of the non-parametric SFH shows that $\sim62\%$ of the stellar mass is created in a burst during the most recent $3$ Myr, a fraction that only marginally ($<0.5\%$) depends on the number of lookback time bins adopted as a prior for the SFH (see Section \ref{sect:analysis}). However, it also indicates that $80\%$ of stars were formed in the last $\lesssim 160$ Myr ($2\sigma$), suggesting that sustained star formation could be happening at earlier phases of mass assembly.

The analysis assuming a constant SFH and nebular emission results in an extremely young stellar population age, $\log$(age/yr)$=6.7^{+0.6}_{-0.4}$. Remarkably, the corresponding best fit SED and main stellar population parameters  are very close ($\Delta\log M_\star \sim0.08$dex, $\Delta A_V\sim 0.02$ mag) to those obtained with the non-parametric SFH,  increasing the overall confidence on this result. Interestingly, the $M_T$ for the non-parametric SFH is $\sim0.2$ dex larger than that for the CSFH, consistent with recent results (e.g., \citealt{tacchella2022}). Indeed, we find a $\sim1$ dex larger value of log(t$_{\ell,80}$/yr) for the non-parametric SFH, consistent with a scenario of considerable star formation at the beginning of mass assembly. Instead, different \textsc{mfrac} values ($0.78$ and $1.0$ for the non-parametric and CSFH, respectively) mitigate the $0.2$ dex systematic difference, leading to similar $M_\star$ estimates. \\

The measured blue $[3.6] - [4.5] \lesssim 0$ mag color might be seen as suggestive of a Balmer jump SED, seen in very young stellar populations (e.g., \citealt{bica1994, inoue2011, schaerer2009, schaerer2010}). However, the $[3.6]-[4.5] < 0$ mag color isn't even significant at $1\sigma$, and the result can also be fit (reduced $\chi^2\sim0.6$) by a very young stellar population ($\sim10^{6.4}$ yr) with no nebular continuum and a modest amount of dust extinction ($A_V \sim 1.1$ mag, leading to a $\sim1$ dex larger stellar mass). 

ALMA results also provide additional insights in this context. The obscured SFR for XDFJ-3811362435 corresponding to $A_V=1.1$ mag is SFR$_\mathrm{obscured}\approx 6 M_\odot$ yr$^{-1}$\footnote{We assumed the IRX-$\beta$ relation  (where IRX$\equiv L_\mathrm{IR}/L_\mathrm{UV}$ is the infrared excess, with $L_\mathrm{IR}$ and $L_\mathrm{UV}$ the infrared and UV luminosities, respectively - e.g., \citealt{meurer1999}) for a \citet{calzetti2000} attenuation curve. This relationship was shifted to match an unattenuated $\beta=-3$, consistent with the intrinsic UV slope of the unattenuated best-fit SED template, resulting in IRX$\sim10$ for $\beta=-3$. The SFR corresponding to $L_\mathrm{IR}$ was estimated adopting the \citet{kennicutt1998} conversion.)}, significantly below the $4\sigma$ threshold ($\sim51M_\odot$yr$^{-1}$ e.g., \citealt{bouwens2020}, after conversion to a \citealt{salpeter1955} IMF) in the deep ALMA continuum mapping of the XDF region (ASPECS: \citealt{walter2016, decarli2019}; see also \citealt{dunlop2017}). This results in no effective help in removing the degeneracy between the different SED solutions. Similarly, the SFR$_\mathrm{obscured}\approx 90 M_\odot$ yr$^{-1}$ for the median SED lies below the detection threshold ($\approx700 M_\odot$ yr$^{-1}$ at $4\sigma$) of the wide-area coverage of GOODS-ALMA (\citealt{franco2018}), suggesting that sensitivities similar to ASPECS would be required to ascertain the existence of brighter (but rarer) dusty sources. An even shallower coverage exists in the FIR for GOODS-N (e.g., \citealt{dowell2014,liu2018}), limiting further considerations. Instead, and qualitatively, the strong  [\ion{O}{3}]$+$H$\beta$ line emission in star-forming galaxies  (equivalent width EW$\gtrsim500-1000$\AA) at $6\lesssim z \lesssim 8$ (see Section \ref{sect:intro} for references) supports a similar scenario with significant nebular continuum at $z\sim10$, just $\sim 150$ Myr earlier than $z\sim8$, consistent with both the non-parametric and CSFH configurations.

\subsection{Comparison to previous studies}

\begin{figure*}
\centering\includegraphics[width=18cm]{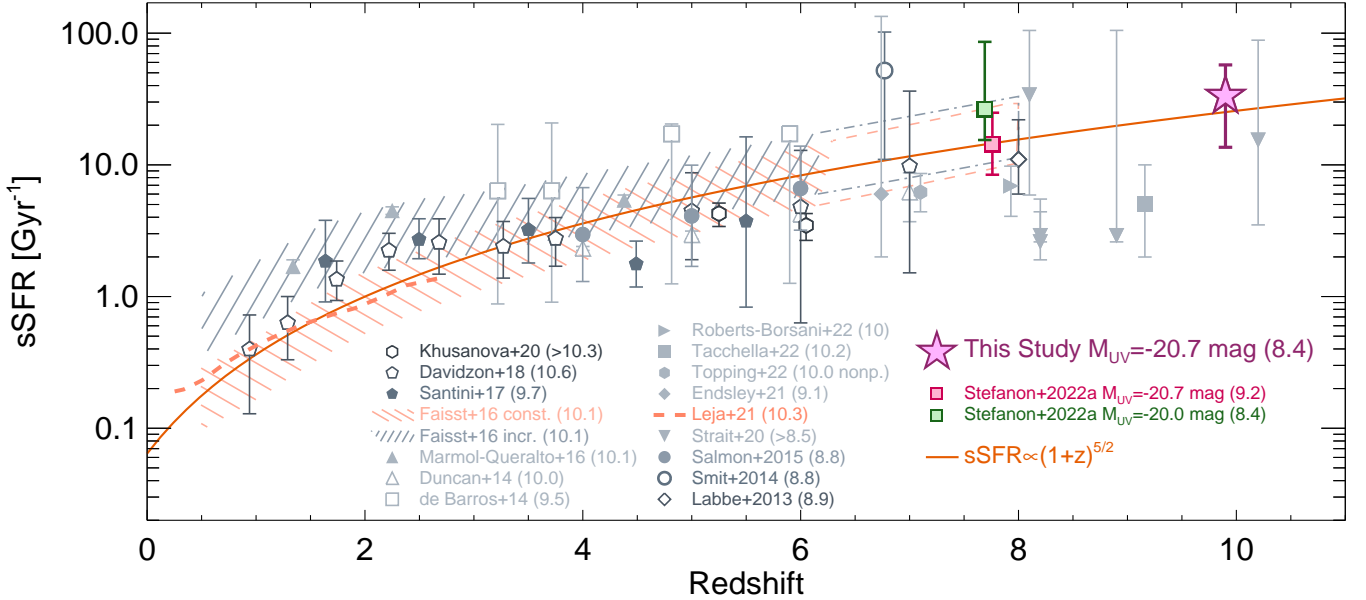}
\caption{Evolution of the sSFR since $z\sim10$. The estimate from this work (filled purple star) is presented together with determinations from the literature indicated by the legend. The number in parenthesis indicates the $\log M_\star$ for each set of measurement, converted to a \citet{salpeter1955} IMF whenever necessary following the prescription of \citet{madau2014}.  The solid orange curve corresponds to the sSFR predicted by the strikingly simple model of \citet{dekel2013}, which builds on the assumption that the formation of stars in galaxies is dominated by the inflow of cold gas driven by the hierarchical merging of the dark matter halos. Our new measurement at $z\sim10$ is consistent with the model prediction, suggesting marginal evolution in the star-formation efficiency of galaxies with cosmic time. This supports previous findings at $z\sim8$ (e.g., \citealt{stefanon2022a}), and is indicative of minimal evolution in the star formation efficiency with cosmic time, beginning as early as just $\sim 500$ Myr after the Big Bang. \label{fig:ssfr}}
\end{figure*}

Our results are consistent with the average blue color and young stellar ages \citet{stefanon2022a} found for star-forming galaxies at $z\sim8$, and with recent analyses of individual sources in samples at $z>7$ (e.g., \citealt{ strait2020,topping2022}). Nonetheless, indication of red IRAC $[3.6]-[4.5]>0$ mag colors, suggestive of the presence of a Balmer break and more evolved stellar populations, have been found for \textit{some} individual sources at similar redshifts (e.g., \citealt{roberts-borsani2016, huang2016, hashimoto2018, strait2020, laporte2021, tacchella2022, roberts-borsani2022}). 

One potential limitation of current results at $z\sim9$ consists in the small number of sources with spectroscopic confirmation (only four galaxies with $8.5\le z_\mathrm{spec}\le 9.5$ exist in the literature: \citealt{zitrin2015, hashimoto2018, laporte2021, larson2022}). The typical uncertainties from photometric redshifts at $z\sim9$ generally do not enable us to ascertain whether the IRAC $4.5\mu$m band is probing exclusively the stellar+nebular continuum or if instead it is also intercepting the contribution from strong emission by [\ion{O}{3}]+H$\beta$, making the interpretation of red $[3.6]-[4.5]>0$ mag colors more uncertain.

Nonetheless, recent works have shown the existence of sources with $z_\mathrm{spec}\gtrsim9$ and red IRAC colors  (\citealt{hashimoto2018, laporte2021}). MACS1149-JD1  (\citealt{zheng2012, hoag2018, hashimoto2018}), at $z_\mathrm{spec}=9.1$ (\citealt{hashimoto2018}), is perhaps the most enigmatic representative of this category. Its exceptionally red $[3.6]-[4.5]\sim0.9$ mag color, indicative of a pronounced Balmer break, clearly contrasts with the $\sim$ blue $[3.6]-[4.5]\lesssim0$ mag color we find in our study (see Figure \ref{fig:SEDs}). It is also worth noting that while spectroscopic redshifts are necessary to accurately establishing which emission lines could be contributing to the flux densities in IRAC bands, they are not always sufficient to robustly segregate evolved from young stellar population ages for $z\lesssim9$ sources (e.g., GN-z10-3, see \citealt{laporte2021} and Figure 2 of \citealt{topping2022}). The medium-band JWST/NIRCam filters and JWST/NIRSpec observations will soon enable to probe the rest-frame optical continuum in between the emission lines at high-z, providing key information to explore the ages and star-formation histories of the earliest generations of galaxies.

\subsection{Evolution of the sSFR}

The value we obtained for the sSFR is  displayed in Figure \ref{fig:ssfr} together with previous determinations from the literature at $1\lesssim z \lesssim10$ (\citealt{labbe2013, smit2014, duncan2014, salmon2015,marmol2016, santini2017, davidzon2018, khusanova2020,strait2020,endsley2021,faisst2016,leja2021, topping2022, tacchella2022,roberts-borsani2022}).

Our new determination is consistent with the measurements of \citet{stefanon2022a} and \citet{labbe2013} at $z\sim8$, and at $\lesssim 2\sigma$ with those of \citet{strait2020} at $z\gtrsim 8$ and with the median for a sample at $z\sim9$ analized by \citet{tacchella2022}. Comparison of these measurements to those at lower redshifts suggests a sSFR monotonically growing with increasing redshift up to $z\sim10$, albeit with a large $\approx 1$ dex scatter. 

Finally, to gather insights into the relationship between the hierarchical assembly of the dark matter halos and the stellar mass accretion of galaxies, in Figure \ref{fig:ssfr} we also compare the measurements to the evolution of the sSFR predicted by \citet{dekel2013}. This  was obtained through analytical considerations on the Extended Press-Schechter formalism which describes the assembly rate of the dark matter halos (see also \citealt{neistein2008b, weinmann2011, genel2014}). The final values for the sSFR were obtained assuming that the baryonic accretion involved exclusively cold gas, corresponding to a time-independent conversion factor between the specific accretion rate of the dark matter halos and the sSFR.  The agreement between our $z\sim10$ estimate and the model prediction is quite remarkable, considering the minimal set of assumptions adopted in \citet{dekel2013} model.\\

\citet{stefanon2022a} found that galaxy assembly in the early Universe was dominated by the accretion of dark matter halos and cold gas, with minimal evolution in the star formation efficiency (SFE). The determination of the sSFR to $z\sim10$ from the present study extends that earlier result of an unevolving SFE by $\sim 150$ Myr  to just $\sim500$ Myr of cosmic time.  In particular, this result qualitatively supports the non-evolving stellar-to-halo mass ratio at $4\lesssim z \lesssim 10$ inferred by \citet{stefanon2021a} from the analysis of the stellar mass function of star-forming galaxies. This built on the work of \citealt{stefanon2017b} using the evolution of the rest-frame optical luminosity function, and of \citet{bouwens2015, bouwens2021} based on the evolution in the UV LF from $z\sim10$ to $z\sim0$. Interestingly, this is consistent with predictions of recent models (e.g., \citealt{tacchella2018, park2019}).

\section{Conclusions}
\label{sect:conclusions}

We derive the median SED of star-forming galaxies at $z \sim10$, obtained by stacking the Spitzer/IRAC $3.6$ and $4.5\mu$m image stamps, and the \textit{HST} photometry, of four robust $J_{125}$ dropouts previously identified over the GOODS fields. 

Crucial for our study are the deepest data at $3-5\mu$m before JWST starts its operations, provided by the GREATS program (PI: Labb\'e - \citealt{stefanon2021b}), which at these redshifts bracket the Balmer/4000\AA\ break, a proxy for the age of the stellar population.

The most fascinating feature seen in the stacked SED is a blue $[3.6]-[4.5]=-0.18\pm0.25$ mag color. At these redshifts, the $3.6\mu$m band probes the rest-frame UV, while the $4.5\mu$m band intercepts the rest-frame optical light. Given the wavelength contiguity of these two bands, the observed blue IRAC color suggests very young stellar populations. Indeed, a quantitative analysis through SED template fitting performed with \textsc{Prospector} (\citealt{leja2019a,johnson2021}) indicates that 80\% of the stars assembled in the most recent $\lesssim160$ Myr ($2\sigma$). The observed blue color contrasts with the red $[3.6]-[4.5]$ measurements for some sources in recent studies (e.g., MACS1149-JD1 - \citealt{hashimoto2018}).

Finally, using the results from the SED fitting, we compute the sSFR. Our value is generally consistent with previous determinations at $z\lesssim 9$, although some estimates form the literature are only consistent at $\sim2\sigma$. Overall, our analysis confirms the trend for a monotonic increase of the sSFR with increasing redshifts, previously found for $z\lesssim8$ (e.g., \citealt{faisst2016, stefanon2022a}). The sSFR measurements for $3\lesssim z \lesssim10$ are also broadly consistent with the toy-model of \citet{dekel2013}, in which the formation of stars is dominated by the inflow of cold gas driven by the hierarchical merging of the dark matter halos.

This result, at $z\sim10$, corresponding to just $\sim500$ Myr of cosmic time, extends earlier results at $z\sim8$ by \citet{stefanon2022a}, and indicates that galaxy assembly in the early Universe was dominated by the accretion of dark matter halos and cold gas, with minimal evolution in the star formation efficiency (SFE).

Clearly, JWST science operations will provide completely game-changing insights into the stellar populations of galaxies and their prevalence earlier than 500 Myr of cosmic time thanks to the exquisite capabilities of JWST/NIRSpec, NIRCam and MIRI, facilitating a major leap in our understanding of galaxy formation. Our Hubble/Spitzer driven result hints at exciting future insights from JWST's study of the earliest galaxies.

\begin{acknowledgments} 
MS acknowledges support from the CIDEGENT grant CIDEGENT/2021/059. MS and RJB acknowledge support from TOP grant TOP1.16.057. PAO acknowledges support from the Swiss National Science Foundation through the SNSF Professorship grant 190079 `Galaxy Build-up at Cosmic Dawn'. The Cosmic Dawn Center (DAWN) is funded by the Danish National Research Foundation under grant No.\ 140.  We also acknowledge the support of NASA grants HSTAR-13252, HST-GO-13872, HST-GO-13792, and NWO grant 600.065.140.11N211 (vrij competitie). GDI acknowledges support for GREATS under RSA No. 1525754. This paper utilizes observations obtained with the NASA/ESA \textit{Hubble Space Telescope}, retrieved from the Mikulski Archive for Space Telescopes (MAST) at the Space Telescope Science Institute (STScI). STScI is operated by the Association of Universities for Research in Astronomy, Inc. under NASA contract NAS 5-26555. This work is based [in part] on observations made with the \textit{Spitzer Space Telescope}, which was operated by the Jet Propulsion Laboratory, California Institute of Technology under a contract with NASA. Support for this work was provided by NASA through an award issued by JPL/Caltech.
\end{acknowledgments}

\bibliographystyle{apj}

%\bibliography{mybib}

\end{document}